\documentclass[aps,prl,preprint,groupedaddress]{revtex4}
\usepackage{graphicx}

\begin{document}


\title{Integral Coalescence Conditions In  $D \geq 2$ Dimension Space}


\author{Xiao-Yin Pan and Viraht Sahni }
\affiliation{Department of Physics, Brooklyn College and the
Graduate School of the City University of New York,365 Fifth Avenue, New York, New
York 10016. }


\date{\today}

\begin{abstract}
We have derived the integral form of the cusp and node coalescence conditions satisfied by the wavefunction at the coalescence of two charged particles in $D \geq 2$ dimension space. From it we have obtained the differential form of the coalescence conditions.  These expressions reduce to the well-known integral and differential coalescence conditions in $D = 3$ space.  It follows from the results derived that the approximate Laughlin wavefunction for the fractional Quantum Hall Effect satisfies the node coalescence condition.  It is further noted that the integral form makes evident that unlike the electron-nucleus coalescence condition, the differential form of the electron-electron coalescence condition cannot be expressed in terms of the electron density at the point of coalescence. From the integral form, the integral and differential coalescence conditions for the pair-correlation function in $D\geq 2$ dimension space are also derived.  The known differential form of the pair function cusp condition for the uniform electron gas in dimensions $D = 2,3$ constitute a special case of the result derived.    
\end{abstract}

\pacs{}

\maketitle

\section{I. Introduction}

The non relativistic wavefunction exhibits a cusp or a node at the coalescence of any two charged particles because of the singularity associated with the Coulomb interaction between them. The differential form of the cusp coalescence condition satisfied by the wavefunction for both electron-electron and electron-nucleus coalescence was derived for dimensions $D = 3$ by Kato \cite{1}.  This form of the cusp condition is in terms of the spherical average of the wavefunction taken about the singularity.  It is this differential form that is most extensively employed in the literature.  The integral form of the electron-electron and electron-nucleus cusp and node coalescence condition for $D = 3$ was originally conjectured  by Bingel\cite{2}.  They have been derived rigorously from the Schrodinger equation by Pack and Byers Brown \cite{3}.  The integral form of the coalescence conditions, however, is more general, and therefore more significant, because it retains the angular dependence of the wavefunction at coalescence.  The differential form is then readily obtained from the integral form by spherically averaging and differentiating.  An example of the usefulness of the integral form is the following.  Over the past two and a half decades, there has been uncertainty \cite{4} about whether the local effective potential energy of Kohn-Sham density functional theory \cite{5} was finite or singular at the nucleus of atoms.  Employing the differential form of the cusp condition, it was initially proved \cite{4} that the potential energy is finite at the nucleus of spherically symmetric atoms.  However, using the integral form, the conclusion of finiteness can be generalized \cite{6} to atoms of arbitrary symmetry, to molecules, and to periodic and aperiodic solids.  As another example, it is well known that the differential form of the electron-nucleus cusp condition may be expressed \cite{7} in terms of the spherical average of the electron density about the nucleus.  It has consequently been assumed \cite{8} that the differential form of the electron-electron cusp condition can similarly be expressed in terms of the density at the point of electron coalescence.  The integral form of the electron-electron cusp condition makes clear that no such expression in terms of the density can exist.\\

 At present,  there is considerable interest in both lower dimensional systems such as the two-dimensional $(D = 2)$ electron gas \cite{9} , the  two-dimensional gas in the presence of a magnetic field ( the Quantum Hall Effect) \cite{10}, and in their higher-dimensional $D\geq 4$ generalizations \cite{11}.   Motivated by this interest, we have derived the integral form of the cusp and node coalescence condition for dimensions $D \geq 2$. It becomes evident that the approximate Laughlin\cite{10} wavefunction for the fractional Quantum Hall effect satisfies the  $D = 2$ node coalescence condition.  The $D\geq 2$     differential form of the cusp condition is then obtained by spherically averaging about the singularity over D-dimensional space and differentiating.  Employing the integral cusp condition, we have also derived the integral and differential forms of the pair correlation function at the coalescence of two identical particles in $D \geq 2$ dimensions. The differential form of the cusp condition for the pair correlation function for the uniform electron gas for $D=3$ \cite{12} and $D=2$\cite{13} constitute a special case of our result.
 The use of cusp conditions for the pair-correlation function in $D=3$ spin-density functional theory is expounded in Ref. 14. More recent work on nuclear cusp conditions relevant to density functional theory is described in Ref.15.\\

\section{II . DERIVATION}

Consider a nonrelativistic system of $N$ charged particles in D $(\geq 2)$ dimension space with the Hamitonian ($\hbar=e=1$)
\begin{equation}
 {\hat H}= -\sum_{i=1}^{N} \frac{1}{2 m_{i}} \nabla_{i}^{2} +\sum_{j>i=1}^{N} \frac{Z_{i} Z_{j}}{r_{ij}}, 
\end{equation}   
where $m_{i}$ and $Z_{i}$ are the mass and charge of the ith particle, and $r_{ij}=|{\bf r}_{i}-{\bf r}_{j}|$. In D dimesion space, ${\bf r}=(x_{1},x_{2},...x_{D})$, $r=\sqrt{\sum_{k=1}^{D} x_{k} x_{k}}$ and $\nabla^{2}=\sum_{k=1}^{D} \frac{\partial}{\partial x_{k}} \frac{\partial}{\partial x_{k}}$. 
Due to the Coulomb potential energy term, the Hamiltonian is singular when any two particle
i and j coalesce($r_{ij}\rightarrow 0$). For  the wavefunction $\Psi ({\bf r}_{1},{\bf r}_{2},...{\bf r}_{N})$ which satisfies the Schr{\" o}dinger equation
 \begin{equation}
 {\hat H} \Psi ({\bf r}_{1},{\bf r}_{2},...{\bf r}_{N})=E \Psi ({\bf r}_{1},{\bf r}_{2},...{\bf r}_{N}),
 \end{equation} 
to be bounded and remain finite at the singularites, it must satisfy a cusp coalescence condition. If the wavefunction vanishes at the singularity, it must satisfy a node coalescence condition.\\

 We are interested in the form 
  of the wavefunction when two particles approach each other, i.e. when $r_{ij}$ is very small. Following Pack and Byers Brown\cite{3} we focus our attention on two particles 1 and 2, and transform their coordinates ${\bf r}_{1} $ and  ${\bf r}_{2} $ to the center of mass ${\bf R}_{12}$
  and relative coordinates ${\bf r}_{12}$ as 
   \begin{eqnarray}
   {\bf R}_{12} & = & \frac{m_{1} {\bf r}_{1}+m_{2} {\bf r}_{2}}{m_{1}+m_{2}}, \nonumber \\
   {\bf r}_{12} & = & {\bf r}_{1}-{\bf r}_{2}.
   \end{eqnarray}
 The Hamiltonian of Eq.(1) may then be rewritten as
  
\begin{eqnarray}
 {\hat H} = -\frac{1}{2 \mu_{12}} \nabla_{{\bf r}_{12}}^{2} + \frac{Z_{1} Z_{2}}{r_{12}}-\frac{1}{2 (m_{1}+m_{2})} \nabla_{{\bf R}_{12}}^{2}+ \sum_{i=3}^{N} \{
 Z_{i} (\frac{Z_{1}}{r_{1i}}+ \frac{Z_{2}}{r_{2i}})\} \nonumber \\
 +\sum_{i=3}^{N} \frac{1}{2 m_{i}} \nabla_{{\bf r}_{i}}^{2}+ \sum_{j>i=3}^{N} \frac{Z_{i} Z_{j}}{r_{ij}},
 \end{eqnarray} 
where $\mu_{12}=\frac{m_{1}m_{2}}{m_{1}+m_{2}}$ is the reduced mass of particles 1 and 2.
When particle $1$ and $2$ are within a small distance of each other( $0<r_{12}<\epsilon$), and all other particles are well separated,  there is only one
singularity in the Hamiltonian. Retaining  only terms of  lower order in  $r_{12}$, Eq. (3) reduces to
\begin{equation}
[ -\frac{1}{2 \mu_{12}} \nabla_{{\bf r}_{12}}^{2} + \frac{Z_{1} Z_{2}}{r_{12}}+ O(\epsilon ^{0})] \Psi ({\bf r}_{1},{\bf r}_{2},...{\bf r}_{N})=0,
\end{equation}  
where $O(\epsilon ^{0})$ refers to  terms of order zero (constant) and higher order in $r_{12}$ and the vector components $({\bf r}_{12})_{k},k=1,2,...D $).\\

 Eq.(4) is a one-electron-atom equation in D dimension space. Furthermore, it is not an eigenvalue equation. We are interested in the solution of the equation that are  finite and continuous at the singularities.  Thus, in the limit as  ${\bf r}_{1}\rightarrow {\bf r}_{2}$, we could write the wavefunction in two parts as
\begin{equation}
  \Psi ({\bf r}_{1},{\bf r}_{2},...{\bf r}_{N})= \Psi ({\bf r}_{2},{\bf r}_{2},{\bf r}_{3}...,{\bf r}_{N})+ \delta \Psi ({\bf r}_{1},{\bf r}_{2},...{\bf r}_{N}),
\end{equation}
where the term $\delta \Psi ({\bf r}_{1},{\bf r}_{2},...{\bf r}_{N})$ must  vanish at the singularity ${\bf r}_{1}={\bf r}_{2}$. From the above it follows that we
need consider only terms of first-order in $r_{12}$ in $\delta \Psi$ when ${\bf r}_{1}$ is near ${\bf r}_{2}$. Dropping the subscript $12$, and writing $r_{12}$ as r, we see that  $\delta \Psi$ could have terms of the form 
  $r_{D} B({\bf r}_{2},{\bf r}_{3}...,{\bf r}_{N})$, or $\sum_{l} r_{l} B_{l}({\bf r}_{2},...{\bf r}_{N})$ where $l$ denotes the $2$ to $(D-1)$-dimensional subspace and 
  $r_{l}$ the distance in the subspace, or ${\bf r}\cdot {\bf C}({\bf r}_{2},...{\bf r}_{N})$. Here, $r_{D}=r$ is just the conventional distance in D dimensional space. As an 
  example of the second kind of terms, consider the case of D$=4$. There are four $3$-dimensional subspaces, and six $2$-dimensional subspaces. The distances in the $3$-dimensional subspaces constitute the terms $\sqrt{x_{1}^{2}+x_{2}^{2}+x_{3}^{2}}$, 
  $\sqrt{x_{1}^{2}+x_{2}^{2}+x_{4}^{2}}$,$\sqrt{x_{2}^{2}+x_{3}^{2}+x_{4}^{2}}$,  and $\sqrt{x_{1}^{2}+x_{4}^{2}+x_{3}^{2}}$. The distances in the $2$-dimensional subspaces 
  are $\sqrt{x_{i}^{2}+x_{j}^{2}}$, where $i\neq j$,and  $i,j=1,2,3,4$.
   In the limit $r\rightarrow 0$, such terms do not affect the O(D) symmetry of the first two terms on the  left side of Eq.(5).  However, the Laplacian of these terms is singular since
 $ \nabla^{2} r_{d} = \frac{d-1}{r}$, where $r_{d}$ is the distance in the d-dimensional subspace. These singularities cannot be cancelled by the 
  Coulomb potential energy singularity term of Eq.(5). Thus,  their coefficients must vanish.
  Therefore, (recovering the subscript $12$), Eq.(6) must be written as
 \begin{equation}
  \Psi ({\bf r}_{1},{\bf r}_{2},...{\bf r}_{N})= \Psi ({\bf r}_{2},{\bf r}_{2},{\bf r}_{3}...,{\bf r}_{N})+ r_{12}  B({\bf r}_{2},{\bf r}_{3},...,{\bf r}_{N}) + {\bf r}_{12} \cdot {\bf C}({\bf r}_{2},{\bf r}_{3},...,{\bf r}_{N})+ O(\epsilon ^{2})
\end{equation}
 Next substitute Eq. (6) for $ \Psi ({\bf r}_{1},{\bf r}_{2},...{\bf r}_{N})$ into Eq.(4) and employ   $\nabla^{2} {\bf r}=0$. For the Coulomb potential energy singularity to be cancelled we have
 \begin{equation}
  -\frac{1}{2 \mu_{12}}  \; \frac{D-1}{r_{12}} B({\bf r}_{2},{\bf r}_{3},...,{\bf r}_{N}) + \frac{Z_{1} Z_{2}} {r_{12}} \Psi ({\bf r}_{2},{\bf r}_{2},{\bf r}_{3},...,{\bf r}_{N})=0.
 \end{equation}
  At the point of coalescence, the wavefunction is $\Psi ({\bf r}_{2},{\bf r}_{2},{\bf r}_{3},...,{\bf r}_{N})$. Since we require the wavefunction to be finite at this point we have
  \begin{equation}
  B({\bf r}_{2},{\bf r}_{3},...,{\bf r}_{N})=\frac{2 Z_{1} Z_{2}\mu_{12}}{D-1} \Psi ({\bf r}_{2},{\bf r}_{2},{\bf r}_{3}...,{\bf r}_{N}).
  \end{equation}  
  Thus, in the limit as $r_{12}\rightarrow 0$, we may write the wavefunction as  
 \begin{equation}
   \Psi ({\bf r}_{1},{\bf r}_{2},...{\bf r}_{N})= \Psi ({\bf r}_{2},{\bf r}_{2},{\bf r}_{3},...,{\bf r}_{N})(1+\frac{2 Z_{1} Z_{2}\mu_{12}}{D-1} r_{12} )+ {\bf r}_{12}\cdot {\bf C}({\bf r}_{2},{\bf r}_{3},...,{\bf r}_{N}).
 \end{equation}  
 This is the general form of the integral \textit{cusp coalesence condition} in D-dimension space.Note that this expression is equally  valid even if the wavefunction vanishes at the point of 
 coalescence, i.e. if   $\Psi ({\bf r}_{2},{\bf r}_{2},...,{\bf r}_{N}) = 0$.  This latter is referred to as a   \textit{node coalescence condition}.
 By taking the spherical average of Eq.(10) about the point of  coalescence in D-dimension space, we obtain the differential form of the cusp condition: 
 \begin{equation}
 (\frac{ \partial \bar{\Psi}}{\partial r_{12}}) |_{r_{12}\rightarrow 0} = \frac{2 Z_{1} Z_{2}\mu_{12}}{D-1} \;\; \Psi(r_{12}=0),
\end{equation} 
 where $\bar{\Psi}$ is the spherically averaged wavefunction. 
  For the electron-nucleus coalescence,
 $Z_{1}=-1, Z_{2}=Z$ the nuclear charge , and  $\mu_{12}\approx m_{e} $ the mass of the electron.  For the electron-electron coalescence,
   $Z_{1}=-1, Z_{2}=-1,  \mu_{12}= \frac{1}{ 2} m_{e} $. In $D=3$ dimensions, the traditional
   integral and differential cusp conditions are recovered.\\
   
   It follows from the definition of the density $\rho({\bf r}_{1})$ which is  
\begin{equation}
\rho({\bf r}_{1})=N \int \Psi^{*}({\bf r}_{1},{\bf r}_{2},...{\bf r}_{N})\Psi ({\bf r}_{1},{\bf r}_{2},...{\bf r}_{N}) d{\bf r}_{2}...d{\bf r}_{N},
\end{equation} 
  that the differential form of the \textit{electron-nucleus} cusp condition may be expessed
  in terms of the density at and about the nucleus. In this case with the nuclear positions fixed, the wavefunction does not depend upon the nuclear coordinates. Then when the electron at ${\bf r}_{1}$ approaches a nucleus, say fixed at the origin, the integral for the density which is over all the remaining $N-1$ electrons can be performed. It is evident from the integral  cusp coalesence condition Eq.(10) and the above definition of the density, that there
  can be no such differential form of the cusp condition for \textit{electron-electron} coalescence. Such expressions in the literature \cite{8} are therefore invalid.\\
   
  It is however, possible to derive the integral and differential forms of the
  cusp condition in D-dimensions for the two-particle correlation function 
  $p({\bf r}_{1} {\bf r}_{2})$ defined as
  \begin{equation}
    p({\bf r}_{1} {\bf r}_{2}) =\frac{N (N-1)}{2} \int \Psi^{*}({\bf r}_{1},{\bf r}_{2},...{\bf r}_{N})\Psi ({\bf r}_{1},{\bf r}_{2},...{\bf r}_{N}) d{\bf r}_{3},...d{\bf r}_{N}.
  \end{equation} 
 In the limit as ${\bf r}_{1}\rightarrow {\bf r}_{2}$ and employing
 the integral cusp expression of Eq.(10) we have  that to first-order in $r_{12}$
  \begin{eqnarray}
  p({\bf r}_{1} {\bf r}_{2}) 
  & =& p({\bf r}_{2} {\bf r}_{2})(1+ \frac{2 Z_{1} Z_{2}\mu_{12}}{D-1} r_{12} )^{2}+ \frac{N (N-1)}{2} \int ({\bf r}_{12}\cdot {\bf C}({\bf r}_{2},...{\bf r}_{N}))^{2} d{\bf r}_{3}...d{\bf r}_{N}  \nonumber \\
  & + & N (N-1)(1+ \frac{2 Z_{1} Z_{2}\mu_{12}}{D-1} r_{12} )\int
  ({\bf r}_{12}\cdot {\bf C}({\bf r}_{2},...{\bf r}_{N})) \Psi ({\bf r}_{2},{\bf r}_{2},...{\bf r}_{N})d{\bf r}_{3}...d{\bf r}_{N} \nonumber \\
  & + & O(\epsilon ^{3}).
  \end{eqnarray}
  This is the integral form of the cusp coalescence condition for the pair-correlation function.
  The differential form of the cusp condition for the correlation function is 
 \begin{equation}
  (\frac{ \partial \bar{p}({\bf r}_{1} {\bf r}_{2})}{\partial r_{12}}) |_{r_{12}\rightarrow 0} = \frac{4 Z_{1} Z_{2}\mu_{12}}{D-1} \;\; p({\bf r}_{2} {\bf r}_{2}).
 \end{equation}
 where $\bar{p}$ is the spherical average of the correlation function.
   The differential cusp condition derived for the uniform electron gas in dimensions $D=3$ and $2$ constitute a special case of this general form of the differential cusp condition.\\

\section{III.  CONCLUDING REMARKS}
  
 In this paper we have derived the integral and differential forms of the coalescence condition satisfied by the wavefunction at the coalescence of two charged particles in $D\geq 2$   dimensions.  The corresponding integral and differential cusp conditions for the pair correlation function in $D\geq 2$ dimensional space are thereby derived.
These results are also valid for the case of a more generalized Hamitonian
which includes a term of the form $\sum_{i} v({\bf r}_{i}) $, where $v({\bf r}_{i}) $
is a local external potential. This is because such terms do not contribute any singularities
at the coalescense of the two charged particles. Such an external potential could
effectively arise for example in the presence of an electric or magnetic field. 
The coalescence conditions constitute rigorous constraints on the construction and evaluation of approximate wavefunctions.  Thus, having derived the integral coalesence condition for $D=2$,
we now understand that the approximate Laughlin wavefunction \cite{10} does in fact satisfy
the \textit{node  coalescence condition}.  The Laughlin wavefunction is
\begin{equation}
\Psi_{m}(z_{1},...,z_{N})=\prod^{N}_{j<k}(z_{j}-z_{k})^{m} \; exp[-\frac{1}{4}\sum^{N}_{l} |z_{l}|^{2}],
\end{equation}
where $m$ is an odd integer, and $z_{j}=x_{j}+i y_{j}$ is the location of the j th
particle expressed as a complex number. It is evident that this wavefunction satisfies
the node coalescence condition for $m \geq 3$. The cusp conditions also impose constraints on the construction of approximate energy functionals within density functional theory \cite{14,16}, and on the wavefunction functionals of the density employed in quantal density functional theory \cite{17}. The integral cusp condition in one dimension, (which is different in form from the result derived for dimension $D\geq 2$), and from which  emanates some very interesting physics, is to be published elsewhere\cite{18}. \\

 This work was supported in part by the Research Foundation of
 CUNY.

\subsection{}
\subsubsection{}


\begin{references}
\bibitem{1}T. Kato, Commun. Pure Appl. Math. \textbf{10}, 151 (1957).
\bibitem{2}W. A. Bingel, Z. Naturforsch. \textbf{18a}, 1249 (1963).
\bibitem{3}R. T. Pack and W. B. Brown, J. Chem. Phys. \textbf{45}, 556 (1966);
 W. A. Bingel, Theoret. Chim. Acta. (Berl.) \textbf{8}, 54 (1967). 
\bibitem{4} Z. Qian and V. Sahni, Int. J. Quantum. Chem. \textbf{79}, 205(2000).
\bibitem{5} W. Kohn abd L. J. Sham, Phys. Rev. \textbf{140},A 1133(1968).

\bibitem{6}X.-Y. Pan and V. Sahni, Phys. Rev. A  \textbf{67}, 012501 (2003).  

\bibitem{7} E. Steiner, J. Chem. Phys. \textbf{39}, 2365 (1963).
\bibitem{8} I. Nagy, Phys. Rev. B  \textbf{52},  1497(1995).  
\bibitem{9} T. Ando, A.B. Fowler, and F. Stern,  Rev. Mod. Phys. \textbf{54}, 437 (1982);
 E. Abrahams, S. V. Kravchenko, and M. P. Sarachik, Rev. Mod. Phys. \textbf{73}, 251 (2001).


\bibitem{10} R. B. Laughlin, Rev. Mod. Phys. \textbf{71},  863 (1998).
\bibitem{11} S. C. Zhang and J. Hu, Science  \textbf{294},  823(2001);
            D. Karabali and V. P. Nair,  Nucl.Phys. B \textbf{641}, 533(2002).
          
            

\bibitem{12} J. C. Kimball, Phys. Rev. A \textbf{7}, 1648 (1973);
A. K. Rajagopal, J. C. Kimball, and M. Banerjee, Phys. Rev. B \textbf{18},
              2339 (1978);
             J. P. Perdew and Y. Wang, Phys. Rev. B \textbf{46}, 12947 (1992).

\bibitem{13}A. K. Rajagopal and J. C. Kimball, Phys. Rev. B \textbf{15}, 2819(1977);
\bibitem{14}A. K. Rajagopal, Adv. in Chemical Physics, \textbf{41}, 59(1980).
\bibitem{15} N. H. March, I.A. Howard, A. Holas, P. Senet, and V. E. Van Doren, Phys. Rev. A
              \textbf{63},  012520 (2000); 
\bibitem{16} P. Gori-Giorgi, C. Attaccalite, S. Moroni, and G. B. Bachelet, cond-mat/0110444(2001).
\bibitem{17} V. Sahni, L. Massa, R. Singh, and M. Slamet, Phys. Rev. Lett, \textbf{87},     113002 (2001); V. Sahni and X.-Y. Pan, Phys. Rev. Lett.  \textbf{90}, 123001 (2003).
\bibitem{18} X.-Y. Pan and V. Sahni, (unpublished).

\end{references}
\end{document}